\documentclass[useAMS,usenatbib]{mn2e}
\usepackage{journal_names}
\usepackage{graphicx,times}
\usepackage{amsmath}
\usepackage[T1]{fontenc}
\usepackage{aecompl}
\usepackage{subfig}
\usepackage{tabularx}
\usepackage{longtable}
\usepackage{pdflscape}
\usepackage{rotating}
\usepackage{multirow}

\def\sun{\hbox{$\odot$}}

\def\la{\mathrel{\hbox{\rlap{\hbox{\lower4pt\hbox{$\sim$}}}\hbox{$<$}}}}
\def\ga{\mathrel{\hbox{\rlap{\hbox{\lower4pt\hbox{$\sim$}}}\hbox{$>$}}}}

\def\deg{{^\circ}}

\newcommand{\HI}{\mbox{\normalsize H\thinspace\footnotesize I}}

\def\apjl   {{ApJL }}
\def\apjs   {{ApJS }}

\title[The Parkes ZOA mass function]{The \protect\mbox{\huge H\thinspace\Large I} mass function in the Parkes \protect\mbox{\huge H\thinspace\Large I} Zone of Avoidance survey}

\author[K.~Said et al.]
{\parbox{\textwidth}{Khaled~Said$^{1}$\thanks{E-mail: khaled.said@anu.edu.au}\thanks{IAU and Gruber Foundation Fellow},
\ Ren\'ee~C.~Kraan-Korteweg$^{2}$,
Lister~Staveley-Smith$^{3,4}$} \vspace{0.4cm}\\
\parbox{\textwidth}{$^{1}$Research School of Astronomy and Astrophysics, Australian National University, Canberra, ACT 2611, Australia\\
$^{2}$ Astronomy Department,  University of Cape Town, Private Bag X3, Rondebosch, 7701, South Africa\\
$^{3}$International Centre for Radio Astronomy Research (ICRAR), M468, The University of Western Australia, 35 Stirling Highway, Crawley, WA 6009, Australia\\
$^{4}$ARC Centre of Excellence for All Sky Astrophysics in 3 Dimensions (ASTRO 3D)}}

\begin{document}

\date{Accepted 2019 April 01. Received 2019 February 05; in original form 2018 September 05}

\pagerange{\pageref{firstpage}--\pageref{lastpage}} \pubyear{2019}

\maketitle

\label{firstpage}

\begin{abstract}
An \HI\ mass function (HIMF) was derived for 751 galaxies selected from the deep Parkes \HI\ survey across the Zone of Avoidance (HIZOA). HIZOA contains both the Great Attractor Wall and the Local Void, two of the most extreme environments in the local Universe, making the sample eminently suitable to explore the overall HIMF as well as its dependence on local environment.  To avoid any selection bias because of the different distances of these large-scale structures, we first used the two-dimensional stepwise maximum-likelihood method for the definition of an average HIMF.  The resulting parameters of a Schechter-type HIMF for the whole sample are $\alpha = -1.33\pm0.05$,  $\log(M_{\rm HI}^*/M_{\odot})=9.93\pm0.04$, and $\phi^* = (3.9\pm0.6)\times 10^{-3}$ Mpc$^{-3}$.  We then used the $k$-th nearest-neighbour method to subdivide the sample into four environments of decreasing local density and derived the Schechter parameters for each subsample. A strong trend is observed, for the slope $\alpha$ of the low-mass end of the HIMF. The slope changes from being nearly flat, i.e. $\alpha = -0.99\pm0.19$ for galaxies residing in the densest bin, to the steep value of  $\alpha = -1.31\pm0.10$ in the lowest density bin. The characteristic mass, however, does not show a clear trend between the highest and lowest density bins. We find similar trends in the low-mass slope when we compare the results for a region dominated by the Great Attractor, and the Local Void, which are found to be over-, respectively underdense by 1.35 and 0.59 compared to the whole sample.
\end{abstract}

\begin{keywords}
galaxies: distances and redshifts -- galaxies: mass function -- cosmology: observations -- cosmology: large-scale structure of Universe
\end{keywords}

\section{Introduction}
The {\HI} mass function (HIMF) is the number of galaxies of a given {\HI} mass in a given volume. The HIMF of galaxies is a key observational component to understand cosmology, large-scale structures of the Universe, and galaxy formation and evolution. It contains important  information about the physical processes that add and remove the {\HI} mass from galaxies, including ram pressure \citep{1972ApJ...176....1G},  thermal evaporation \citep{1977Natur.266..501C}, tidal forces \citep{1984A&A...139...15S}, and gravitational potential \citep{1999ApJ...510L..15B}.

These physical processes usually proceed at different rates in different environments due to the  dependence of interactions between galaxies as a function of the local galaxy density. As such, we do not  expect that HIMF to be universal but that the shape will change from one environment to the other. 

According to the galaxy evolution theory, a number of physical processes occur at a higher rate in high-density environments compared to low-density environments.  One of these processes is the tidal stripping which removes the {\HI} gas from the interacting galaxies \citep{1984A&A...139...15S}.  Another process that is directly proportional to the environmental density is the ram pressure.  It was shown in two sequential papers \citep{1971ApJ...169L..13G,1972ApJ...176....1G} that the hot gas observed in X-rays in the Coma cluster \citep{1971Natur.231..107M} is responsible for the stripping of gas from galaxies within the cluster due to ram pressure. Although tidal and collision stripping are more efficient in groups of galaxies, ram pressure is more efficient in clusters  \citep{2010AJ....139..102E,2013MNRAS.434.3511W}.

A direct way of observing the process of  {\HI} gas removal from galaxies in high-density regions is through dedicated high resolution {\HI} surveys (e.g., \citealt{2009AJ....138.1741C,2012MNRAS.422.1835S,2013MNRAS.428..370S}). Important clues can also be learned from the  hydrodynamical simulations (e.g., \citealt{2005MNRAS.357L..21B}). Such processes can also be detected statistically via constructing the HIMF across different environments ranging from low-density to high-density regions (e.g., \citealt{2001ASPC..240..507V,2002ApJ...567..247R,2005ApJ...621..215S, 2005MNRAS.359L..30Z,2011ApJS..197...28P,2014MNRAS.444.3559M,2016MNRAS.457.4393J,2017MNRAS.472.4832W,2018MNRAS.477....2J}).

Quantitative findings from the latter methods have been a debated topic for over a decade. \cite{2001ASPC..240..507V} performed a VLA blind {\HI} survey of the Ursa Major cluster with the aim to measure the slope of the HIMF down to low masses. They measured a flat slope for this region. \cite{2002ApJ...567..247R} used the Arecibo Dual-Beam Survey to constructed the HIMF for the entire sample and for 38 galaxies within the center of Virgo Cluster. They  concluded that, the HIMF is less steep in high-density regions.  The main limitation for these studies was the small number of galaxies in the low mass bins. 

In 2005, two independent studies used larger number of galaxies to find contradicting conclusions. \cite{2005ApJ...621..215S}  divides their {\HI} galaxy sample into three subsets according to the local density reconstructed via the PSCz survey \citep{1999MNRAS.308....1B}. They found a steeper HIMF  slope in the lowest density sample compared to the other two samples. \cite{2005MNRAS.359L..30Z}, on the other hand, used the $k$-th nearest neighbour metric
\citep{1980ApJ...236..351D} to divide their sample into five sub-samples according to their local density. They found that the low-mass slope of the HIMF  depends only weakly on the local density, being steeper in higher density environments. They found no dependence of the characteristic \HI-mass $M^*$ on the local density.

More recent studies show that this tension holds even within the same survey. \cite{2014MNRAS.444.3559M} and \cite{2016MNRAS.457.4393J} used 40\% and 70\% of the ALFALFA survey \citep{2005AJ....130.2598G}, respectively, and found no significant dependence of the low-mass slope on the local density but a larger characteristic \HI-mass $M^*$ in high-density regions compared with low-density regions. \cite{2018MNRAS.477....2J} later used the complete ALFALFA sample and found larger change in the low-mass slope than previously, in the sense that the low-mass slope is steeper in the high-density environments (not clusters) than in the low-density environment.

In this work, we take advantage of a blind {\HI} survey that is more than twice as sensitive as HIPASS \citep{2004MNRAS.350.1195M,2004AJ....128...16K} to construct a new HIMF. In addition, we probe the HIMF in different environments because this survey covers the most two extreme environment in terms of density in the local Universe, namely the Great Attractor (GA; \citealt{1988ApJ...326...19L,1999A&A...352...39W}) and the Local Void (LV; \citealt{1987ang..book.....T,2008glv..book...13K}). 

This paper is organized as follows: The sample is presented in Section 2. In Section 3, we describe the methodology used to derive the HIMF. The universality of the HIMF is discussed in Section 4. Lastly, we summarize our results in Section 5. Throughout this paper, we used the Hubble constant of $H_0=75$ km s$^{-1}$ Mpc$^{-1}$.  

\section{HIZOA Sample}

This study is based on the deep ``blind'' \HI\ Zone of Avoidance survey (HIZOA) performed with the multibeam receiver of the 64\,m Parkes radio telescope. This survey covers the inner Milky Way, $|b| < 5\deg$, for the Galactic longitude range accessible from Parkes, $196\deg < \ell < 52\deg$. It includes both the southern ZOA \citep{2016AJ....151...52S} and its extension to the north \citep{2005AJ....129..220D}. It was originally designed to unveil the large-scale structures hidden behind the Milky Way. Significant parts of dynamically influential large-scale structures in the local Universe remained obscured from our view due to dust absorption and stellar crowding. This includes the Great Attractor region, considered to be an extreme overdensity in the nearby Universe (\citealt{1988ApJ...326...19L,1999A&A...352...39W}), and the Local Void, which is the largest nearby void (\citealt{1987ang..book.....T,2008glv..book...13K}) and has been mapped particularly poorly due to its centre being located behind the Galactic Bulge. Further notable structures are the Puppis and Ophiuchus clusters \citep[see][for details]{2016AJ....151...52S}. 

Given the dust obscuration and stellar confusion problems, systematic \HI\ surveys are the only method to chart these structures \citep{2016AJ....151...52S} and to quantitatively assess the extent of overdensities and underdensities.  The survey achieved an rms sensitivity of 6 mJy in the velocity range of $v < 12\,000$ km\,s$^{-1}$. Galaxies were detected within the range of $\log M_{\rm HI}/M_{\odot}$ of $6.5 - 11.0$ and, most importantly, the survey was sensitive to galaxies well below the characteristic \HI-mass $M^*_{\rm HI}$ at the distance of the Great Attractor. 

The primary goal of this paper is therefore to derive an average HIMF from HIZOA, and to explore the variation in the HIMF in some of the most extreme environments in the nearby Universe. The results will also improve our knowledge of the over/under-densities of regions obscured by the ZOA. These mass density estimates will be crucial in improving our understanding of the observed cosmic flow fields in the ZOA (e.g., \citealt{2006MNRAS.373...45E,2011MNRAS.416.2840L,2017MNRAS.466L..29K,2017MNRAS.471.3087S})

\section{Derivation of the HIZOA \protect\HI\ mass function}

We derive the {\HI} mass function for all galaxies in the HIZOA survey that meet our selection criteria. The HIZOA survey has been characterized by three completeness limits: (1) velocity-integrated flux, (2) mean flux density, and (3) a hybrid limit (see Table 4 in \citealt{2016AJ....151...52S}). Above these limits lie between 70 per cent and 90 per cent of the HIZOA sample. 
These limits were derived using a modified version of the $V/V_{\rm max}$ test \citep{2001MNRAS.324...51R} which, unlike the traditional $V/V_{\rm max}$ test \citep{1968ApJ...151..393S}, is insensitive to the presence of large-scale structure. For the current study, we use the hybrid limit, which has the sharpest cutoff and therefore most likely to represent the true HIZOA completeness limit.
The maximum volume to which a HIZOA galaxy can be detected (ignoring the profile resolution correction) with certainty is therefore
\begin{equation}
V_{\rm max} = \frac{4\pi f_{\rm sky}}{3} \left(\frac{M_{\rm HI}}{8.815\times10^5 S_{\rm lim} W_{50}^{0.74}}\right)^{3/2} ,
\end{equation}
where $S_{\rm lim} = 18$ mJy represents the corresponding $T=-2$ completeness limit \citep{2016AJ....151...52S}, $W_{50}$ is the velocity width (in km\,s$^{-1}$), $f_{\rm sky}$ is the fraction of the sky covered by the survey, and $M_{\rm HI}$ is given by
\begin{equation}
M_{\rm HI} = 2.356 \times 10^5  D^2  F_{\rm HI} ~{\rm M}_{\sun},
\end{equation}
where $D$ is the distance in Mpc and $F_{\rm HI}$ is the flux integral. We can then fit a Schechter function \citep{1976ApJ...203..297S} to the binned density estimates of the form:
\begin{equation}
\Phi(M_{\rm HI}) = \text{ln}10 \phi^{*} \left(\frac{M_{\rm HI}}{M_{*}}\right)^{\alpha+1} e^{- \frac{M_{\rm HI}}{M_{*}}},
\label{Sch_form}
\end{equation}
to the derived {\HI} mass function. 

However, as with the $V/V_{\rm max}$ test, a major problem of the $1/V_{\rm max}$ method for deriving the HI mass function is sensitivity to large-scale structure, such that the measured mass function does not represent the universal HI mass function. 
For sub-samples where it can be assumed that the HIMF has a similar shape, but different normalisation, the two-dimensional stepwise maximum likelihood method (2DSWML: \citealt{2000MNRAS.312..557L,2003AJ....125.2842Z,2005MNRAS.359L..30Z}) can be deployed.

The 2DSWML method 
finds the maximum likelihood solutions for the number density of galaxies $\phi_{jk}$ as a function of {\HI} mass and velocity width by iterating from an initial guess:
\begin{equation}
\phi_{jk} = \frac{\sum\limits_{i=1}^{N_g} \delta(M_i-Mj,W_i-W_k)}{\sum\limits_{i=1}^{N_g}\frac{H_{ijk} \Delta M \Delta W}{\sum\limits_{l=1}^{N_M} \sum\limits_{m=1}^{N_W} \phi_{lm} H_{ilm} \Delta M \Delta W}},
\label{main2dswml}
\end{equation}
where $\delta_{ijk}$ is a function equals one if a galaxy $i$ belongs to the {\HI} mass bin $j$ and velocity width bin $k$, and equals zero otherwise. The function $H_{ijk}$ equals the fraction available of the bin $jk$ in which a galaxy $i$ resides \citep{2003AJ....125.2842Z}. 

Equation \ref{main2dswml} can be interpreted in different ways. The denominator for example, presents the effective volume available to galaxies in bin $jk$, and can be modified to find the effective volume available to each galaxy as in \cite{2005MNRAS.359L..30Z}. Marginalizing Eq. \ref{main2dswml} over the velocity width $W$ gives the {\HI} mass function \citep{2010ApJ...723.1359M,2016MNRAS.457.4393J} while marginalizing over the {\HI} mass $M_{\rm HI}$ gives the velocity width function \citep{2011ApJ...739...38P}. 

\begin{figure}
\begin{center}
\includegraphics[scale=0.4]{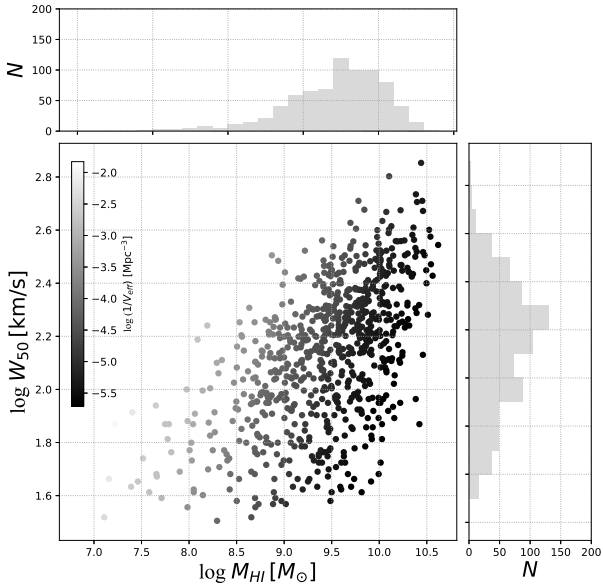}
\caption[The galaxy density distribution in the $M_{\rm HI}-W_{50}$ plane]{The galaxy density distribution in the $M_{\rm HI}-W_{50}$ plane. The gray scale represents the reciprocal of the effective volume for each galaxy in which it can be found. This effective volume is calculated via the two-dimensional stepwise maximum likelihood method.}
\label{2Dhist}
\end{center}
\end{figure}

We used Eq. \ref{main2dswml} to evaluate the effective volume available to each single galaxy after applying the completeness function of the HIZOA survey. Figure \ref{2Dhist} shows the galaxy density distribution in the $M_{\rm HI}-W_{50}$ plane where the gray scale represents the reciprocal of the effective volume available to each galaxy. There is a strong correlation between the linewidth and the effective volume. The histogram of the linewidth on the right-hand side of the plot shows that our sample contains a fair distribution of large and small galaxies, following an approximately Gaussian distribution. 

\begin{figure}
\begin{center}
\includegraphics[scale=0.4]{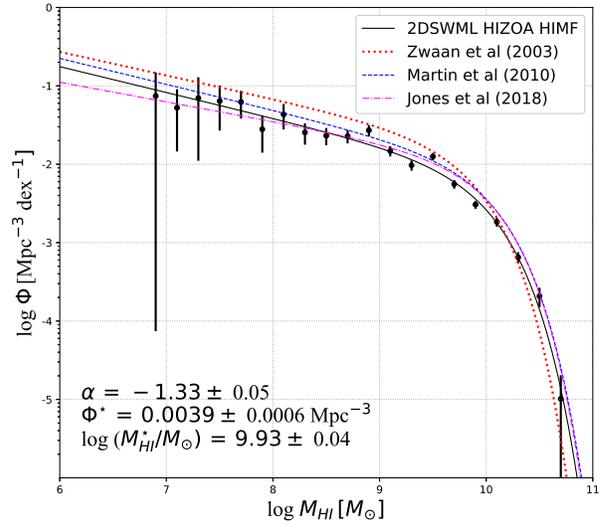}
\caption[HIZOA \protect\HI\ mass function derived via the two-dimensional stepwise maximum likelihood method]{HIZOA {\HI} mass function derived via the two-dimensional stepwise maximum likelihood method (Black-solid line). For comparison, the {\HI} mass function of HIPASS BGC (dotted-red line; \citealt{2003AJ....125.2842Z}),  ALFALFA 40\% (dashed-blue  line; \citealt{2010ApJ...723.1359M}), and ALFALFA 100\% (dashed-dotted-fuchsia line; \citealt{2018MNRAS.477....2J}) are also displayed.}
\label{HIZOA_2dswml_HIMF}
\end{center}
\end{figure}

We then marginalized $\phi_{jk}$ over the velocity width to calculate the HIMF as:
\begin{equation}
\phi_j = \sum\limits_{k=1}^{N_W} \phi_{jk} \Delta W
\end{equation}
Figure \ref{HIZOA_2dswml_HIMF} shows the result of this process for the whole sample above the completeness limit.
The dots present the number space density in {\HI} mass bins, the solid line is the best fit Schechter function of the form presented in Eq. \ref{Sch_form}, and the error bars are the Poisson errors. The final parameters of the {\HI} mass function are:
\begin{itemize}
\item $\alpha = -1.33\pm0.05$
\item $\log(M_{HI}^*/M_{\odot})=9.93\pm0.04$
\item $\phi^* = (3.9\pm0.6)\times 10^{-3}$ Mpc$^{-3}$
\end{itemize}
For comparison, we add to Fig. \ref{HIZOA_2dswml_HIMF} the \cite{2003AJ....125.2842Z} results as a dotted line,  \cite{2010ApJ...723.1359M} as a dashed line, and \cite{2018MNRAS.477....2J} as a dashed-dotted line. The faint-end slope derived here agrees, within the uncertainty, with \cite{2003AJ....125.2842Z}, \cite{2005MNRAS.359L..30Z}, \cite{2010ApJ...723.1359M}, and  \cite{2018MNRAS.477....2J}. The knee of the {\HI} mass function derived here agrees more with the \cite{2010ApJ...723.1359M} and  \cite{2018MNRAS.477....2J} results than \cite{2003AJ....125.2842Z} and \cite{2005MNRAS.359L..30Z}, perhaps because the HIZOA survey is more closely matched in sensitivity to ALFALFA than HIPASS \citep{2016AJ....151...52S}. 

\section{Universality of the HIMF }
In this section, we will extend our analysis and use the HIZOA sample to test the universality of the HIMF, i.e. we will derive the Schechter function parameters ($\alpha$ and $M_{\rm HI}^*$) across various environments of local densities, from extreme low-density to high-density regions. To accomplish this we divide the HIZOA galaxies into sub-samples using two independent methods to infer different environments: (i) the Bayesian nearest neighbours \citep{2005AJ....129.1096I} to determine the local density in which each galaxies resides; and (ii) extreme large-scale structures which, in the case of HIZOA, are the Great Attractor and the Local Void \citep{2016AJ....151...52S}.

\subsection{Bayesian $k$--th nearest neighbours algorithm}
\citet{1980ApJ...236..351D}, showed that the density and the associated error can be estimated by using the distance to the $k$--th nearest neighbour, $d_k$ as $\Sigma_k = {K}/{\frac{4}{3}\pi d_k}$, and $\sigma_{\Sigma_k} = \sqrt{K}/{\frac{4}{3}\pi d_k}$, respectively. Therefore, the fractional error decreases with $k$ at the cost of decreasing the spatial resolution. \citet{2005AJ....129.1096I}, proposed a Bayesian based version of the $k$-th nearest neighbour algorithm to decrease the fractional error without decreasing the spatial resolution. This is achieved by calculating distances to all $k$ nearest neighbours, $d_i$, where $i=1,2,...,k$, instead of calculating the distance to only the $k$-th nearest neighbour, $d_k$, as,

\begin{equation}
\Sigma_k = \frac{C}{\sum\limits_{i=1}^{k}d_{i}^{3}}
\label{k-nearest-neighbour}
\end{equation}
where the normalization C is calculated as,
\begin{equation}
C = \frac{k(k+1)}{2\times\frac{4}{3}\pi}.
\end{equation}

In case of a sparse sample, the above Bayesian method \citep{2005AJ....129.1096I} is less biased compare to the classic $k$-th nearest neighbour method (\citealt{1980ApJ...236..351D}; for comparison, we apply this classic method in Appendix A). The next step in this analysis is to choose the number of neighbours. \cite{1985ApJ...298...80C} showed that the sixth neighbour is a good choice to reduce fluctuations and still retain locality. For that reason we calculate distances to all the six nearest neighbours.

\begin{figure}
\begin{center}
\includegraphics[scale=0.55]{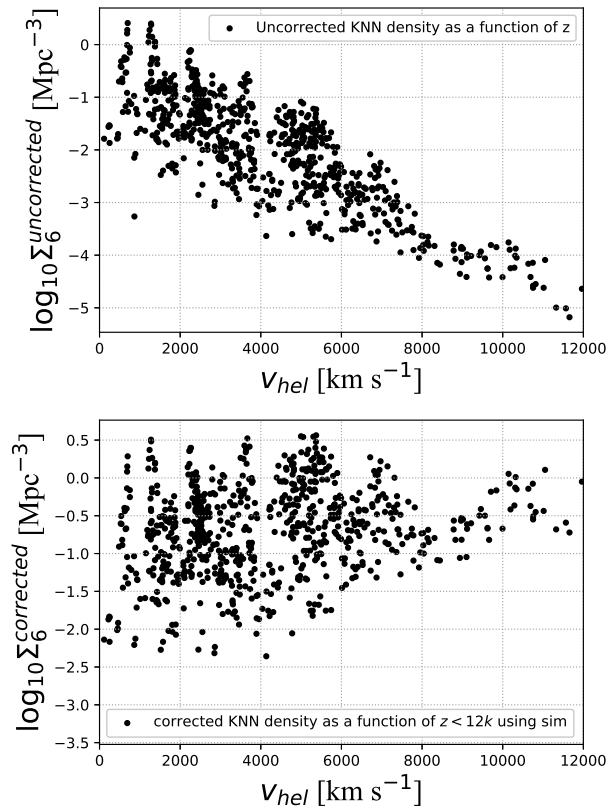}
\caption{Scatter plot of the sixth nearest neighbour density, $\Sigma_6$, using the Bayesian $k$ nearest neighbours algorithm for HIZOA sample. The top-panel shows the uncorrected $\Sigma_6$ while the bottom-panel shows the corrected $\Sigma_6$ using the method presented in Appendix \ref{BBB}}
\label{uncor_vs_cor}
\end{center}
\end{figure}

Another fact that needs to be accounted for is that, for a flux-limited sample, galaxies will be more luminous and appear to be more isolated at larger distances. This implies that a value for $\Sigma_6$ which decreases with distance, even where there is no change in local density. This effect is clearly observed in the top-panel of Fig. \ref{uncor_vs_cor}. We corrected for this effect using flux and volume-limited samples derived from the $S^3-SAX$ semi-analytic simulation. This process is described step-by-step in Appendix \ref{BBB}. The corrected $\Sigma_6$ is shown in the bottom-panel of Fig. \ref{uncor_vs_cor}, and removes significant differences in the distance distribution of samples of different density. The bottom-panel of Fig. \ref{uncor_vs_cor} also shows that our sample becomes very sparse at high-redshift, so we limit all our calculation to galaxies with $v_{hel} \leq 8000$ km s$^{-1}$

\begin{figure}
\begin{center}
\includegraphics[scale=0.35]{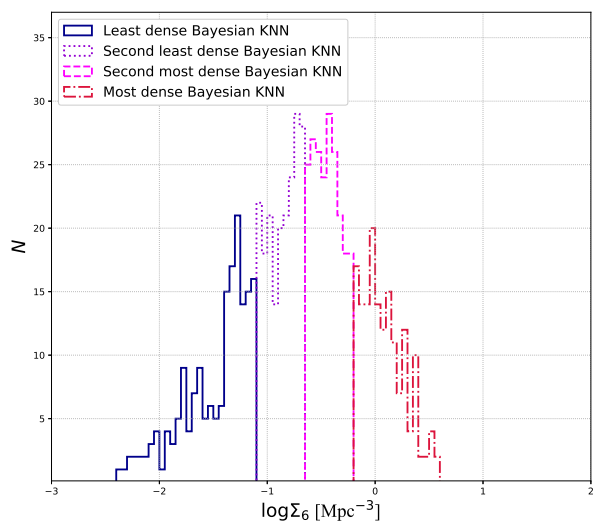}
\caption{Histogram of the sixth nearest neighbour density, $\Sigma_6$, using the Bayesian $k$ nearest neighbours algorithm for the HIZOA galaxies divided into four groups according to $\Sigma_6$.}
\label{density_as_hist_Color}
\end{center}
\end{figure}

\begin{figure}
\begin{center}
\includegraphics[scale=0.52]{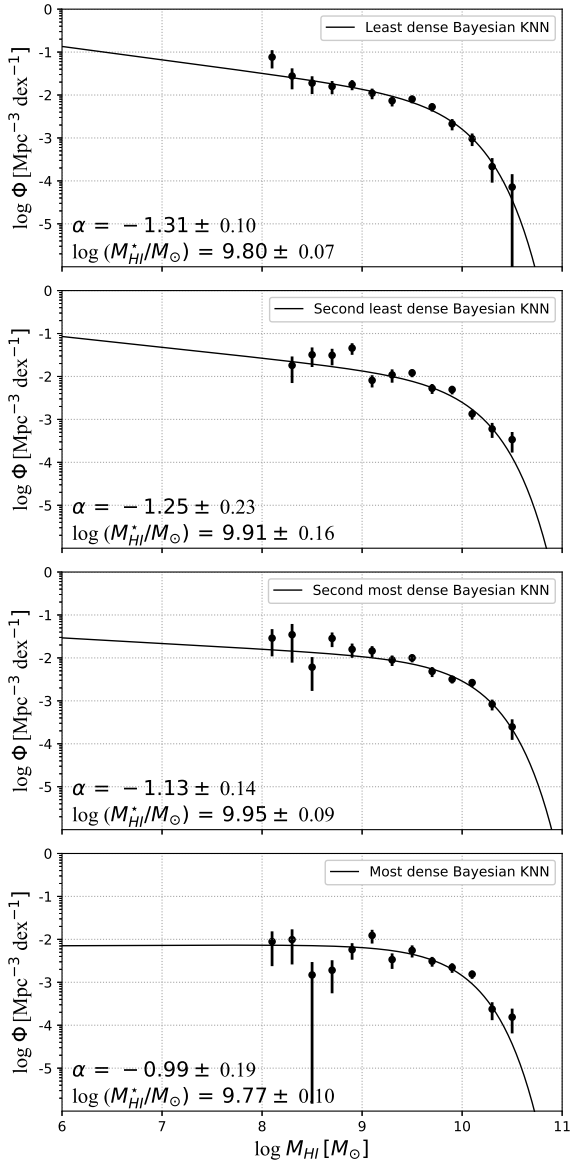}
\caption[Four HIZOA]{Four HIZOA {\HI} mass function derived via the 2DSWML method based on the Bayesian density estimate. For comparison, the HIMFs are offset vertically to the normalization of the whole sample HIMF in Fig. \ref{HIZOA_2dswml_HIMF}.}
\label{compare_density_Alpha_Mstar}
\end{center}
\end{figure}

Figure \ref{density_as_hist_Color} shows a histogram of the corrected sixth nearest neighbour density, $\Sigma_6$.
We divided the whole sample into four sub-samples and plot them on a colour scale from the least dense regions (solid-blue) to the most dense regions (dotted-dashed-red).
For each of these sub-samples we derived the HIMF using the 2DSWML method. Figure \ref{compare_density_Alpha_Mstar} shows the four derived HIMFs ranging from the lowest-density region (top panel) to the highest-density region (bottom panel). For comparison, we use the same mass range of $\log M_{\rm HI}/M_{\odot} =$  $8.0 - 10.5$ for the four sub-samples.




Table \ref{Schechter_P_B} presents the derived Schechter parameters for these four sub-samples ranging from the least dense region to the most dense region based on the Bayesian version of the $k$-th nearest neighbours algorithm.
Although the number of galaxies at the low mass end of each sub-sample are small, we find a clear trend: the HIMF is more shallow in high-density regions than in low-density regions.

\begin{figure}
\begin{center}
\includegraphics[scale=0.52]{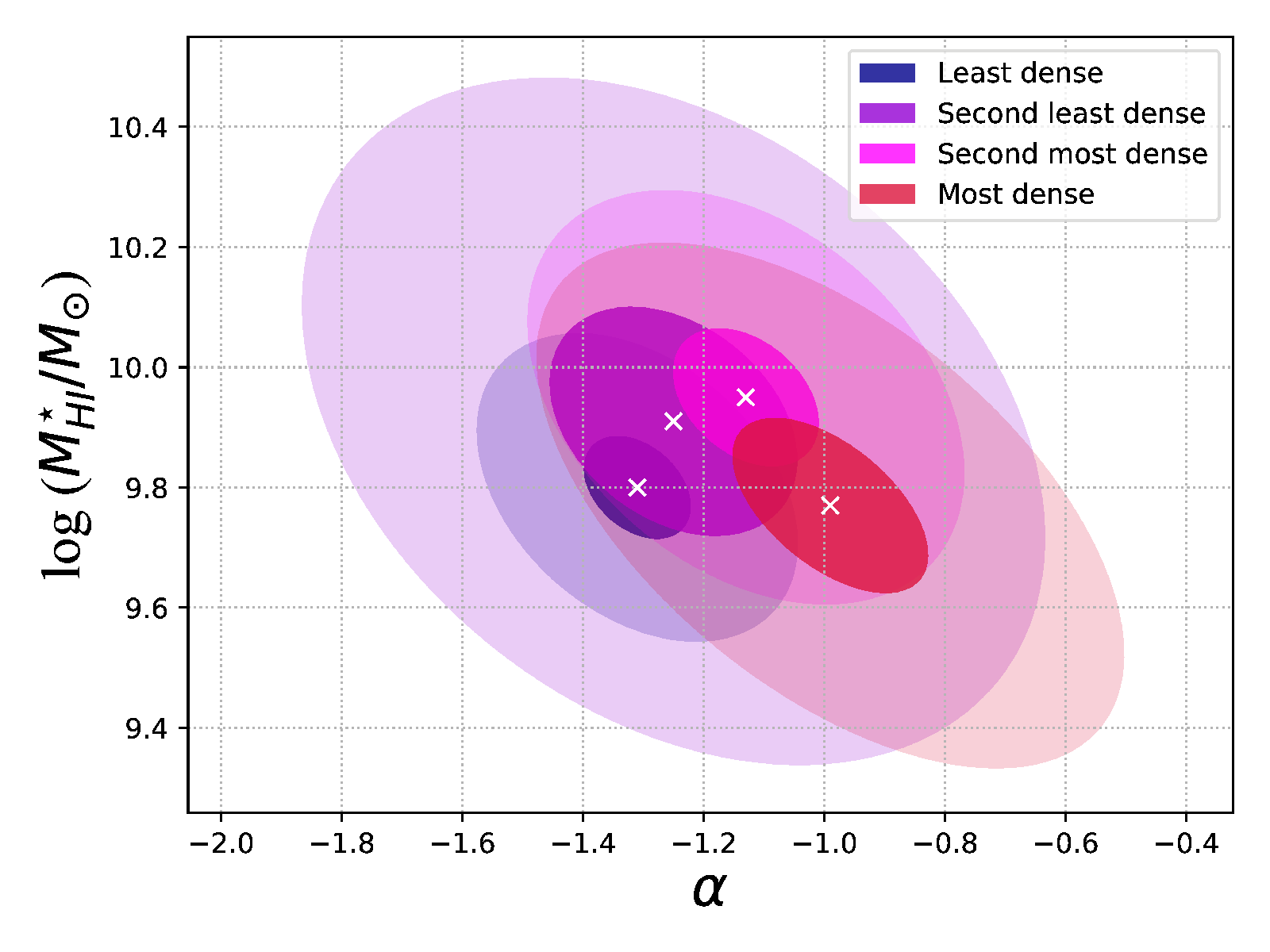}
\caption[Schechter parameters for the four sub-samples divided according to their local density]{Schechter parameters for the four sub-samples divided according to their local density projected on the $\alpha-\log M_{HI}^*$ plane. Ellipses are defined with $(\alpha,\log M_{HI}^*,\sigma_{\alpha},\sigma_{\log M_{HI}^*},\theta)$ and plotted on a colour scale from the least dense regions (blue) to the most dense regions (red). Dark and light ellipses show $1\sigma$ and $3\sigma$ contours, respectively.}
\label{four_regions}
\end{center}
\end{figure}

\begin{table}
\begin{center}
\caption[Schechter parameters for the HIMF]{Schechter parameters for the four sub-samples divided according to their local density derived by the Bayesian version of the $k$-th nearest neighbours algorithm.}
\begin{tabular}{l c c}
\hline
\hline
Region  &   $\alpha$ & $\log(M_{HI}^*/M_{\odot})$\\
\hline
Least dense &   $-1.31\pm0.10$ &  $9.80\pm0.07$  \\
Second least dense &   $-1.25\pm0.23$ &  $9.91\pm0.16$ \\
Second most dense &   $-1.13\pm0.14$ &  $9.95\pm0.09$ \\
Most dense &   $-0.99\pm0.19$ &  $9.77\pm0.10$ \\
\hline
\end{tabular}
\label{Schechter_P_B}
\end{center}
\end{table} 

A projection of the derived Schechter parameters on the $\alpha-\log M_{HI}^*$ plane is shown in Fig. \ref{four_regions}. Confidence ellipses are defined with $(\alpha,\log M_{HI}^*,\sigma_{\alpha},\sigma_{\log M_{HI}^*},\theta)$, where $\theta$ is the angle between the $x$-axis and the major axis of the ellipses, and is measured via the covariance matrix of the correlated parameters:
\begin{equation}
\tan(2\theta) = 2 \frac{\sigma_{\alpha \log M_{HI}^*}}{\sigma^2_{\alpha}-\sigma^2_{\log M_{HI}^*}}.
\end{equation}
For uncorrelated parameters $\theta = \sigma_{xy} = \sigma_{yx} = 0$. This is contrary to what is found in  Fig.~\ref{four_regions} which displays the $1\sigma$  and $3\sigma$ confidence contours (dark and light ellipses respectively) which shows the Schechter parameters to be strongly correlated. 
We therefore find the HIMF not to be universal in shape. The derived parameters for the least dense environment and most dense environment deviate from each other. In particular, the slope $\alpha$ increases with density.

\subsection{Known large-scale structures in  ZOA}

As previously mentioned, the HIZOA survey covers two of the most extreme galaxy structures in the Local Universe with the Great Attractor (GA) and the Local Void (LV). This provides us with two different laboratories of different density environment. We will derive the HIMF for each of these two structures.

The first structure in the HIZOA survey \citep{2016AJ....151...52S} is the GA Wall crossing. This region dominates the HIZOA redshift survey and extends from $v_{\rm LG} \sim 1500$ to $\sim 7500$ km\,s$^{-1}$ and from $l \sim 280^\circ$ to $\sim 330^\circ$. The second prominent structure, the LV extends from $l \sim 330^\circ$ to $\sim 45^\circ$ and spreads out over a volume to $v_{\rm LG} \sim 6000$km\,s$^{-1}$ \citep{1987ang..book.....T,2005AJ....129..220D,2008glv..book...13K}.  The southern part of the LV lies within the main HIZOA survey \citep{2016AJ....151...52S}, whereas the northern part of is covered by the northern extension \citep{2005AJ....129..220D}. 

\begin{figure}
\begin{center}
\includegraphics[scale=0.55]{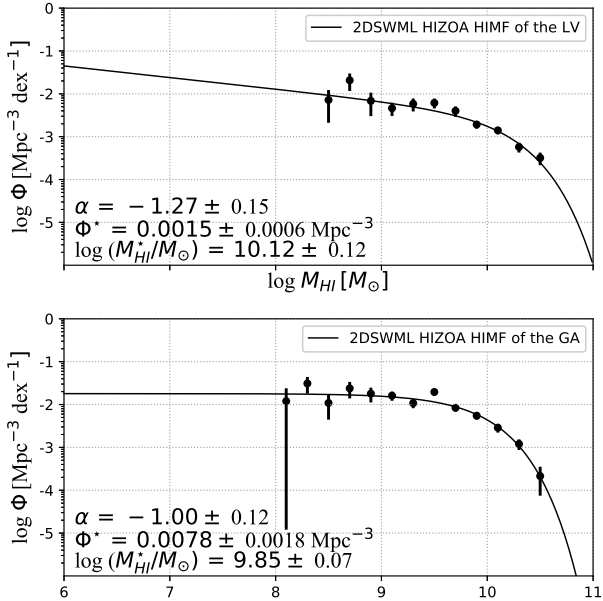}
\caption[HIZOA \protect\HI\ mass function derived via the 2DSWML method]{HIZOA {\HI} mass function derived via the 2DSWML method for LV (top-panel) and GA (bottom-panel) regions.}
\label{HIZOA_2dswml_HIMF_GA}
\end{center}
\end{figure}

Figures \ref{HIZOA_2dswml_HIMF_GA} shows the HIMF derived from the LV (top-panel) and GA (bottom-panel) sub-samples, respectively, and Table \ref{Schechter_P_GA_LV} lists the derived Schechter parameters for the two regions. Again for comparison purposes, we use the same mass range of $\log M_{\rm HI}/M_{\odot} =$  $8.0 - 10.5$ for the two  sub-samples.
There is a trend of increasing $\alpha$ 
when moving from the low-density region (LV) to high-density region (GA). This result is in \textbf{excellent} agreement with the analysis based on the Bayesian $k$-th nearest neighbours.

\begin{table}
\begin{center}
\caption[Schechter parameters for the HIMF]{Schechter parameters for the LV and GA regions.}
\begin{tabular}{l c c c}
\hline
\hline
Region  &   $\alpha$ & $\log(M_{HI}^*/M_{\odot})$ & $\phi^*$ (Mpc$^{-3}$)\\
\hline
LV &   $-1.27\pm0.15$ &  $10.12\pm0.12$  & $0.0015\pm0.0006$\\
GA &   $-1.00\pm0.12$ &  $9.85\pm0.07$ & $0.0078\pm0.0018$ \\
\hline
\end{tabular}
\label{Schechter_P_GA_LV}
\end{center}
\end{table}

\begin{figure}
\begin{center}
\includegraphics[scale=0.52]{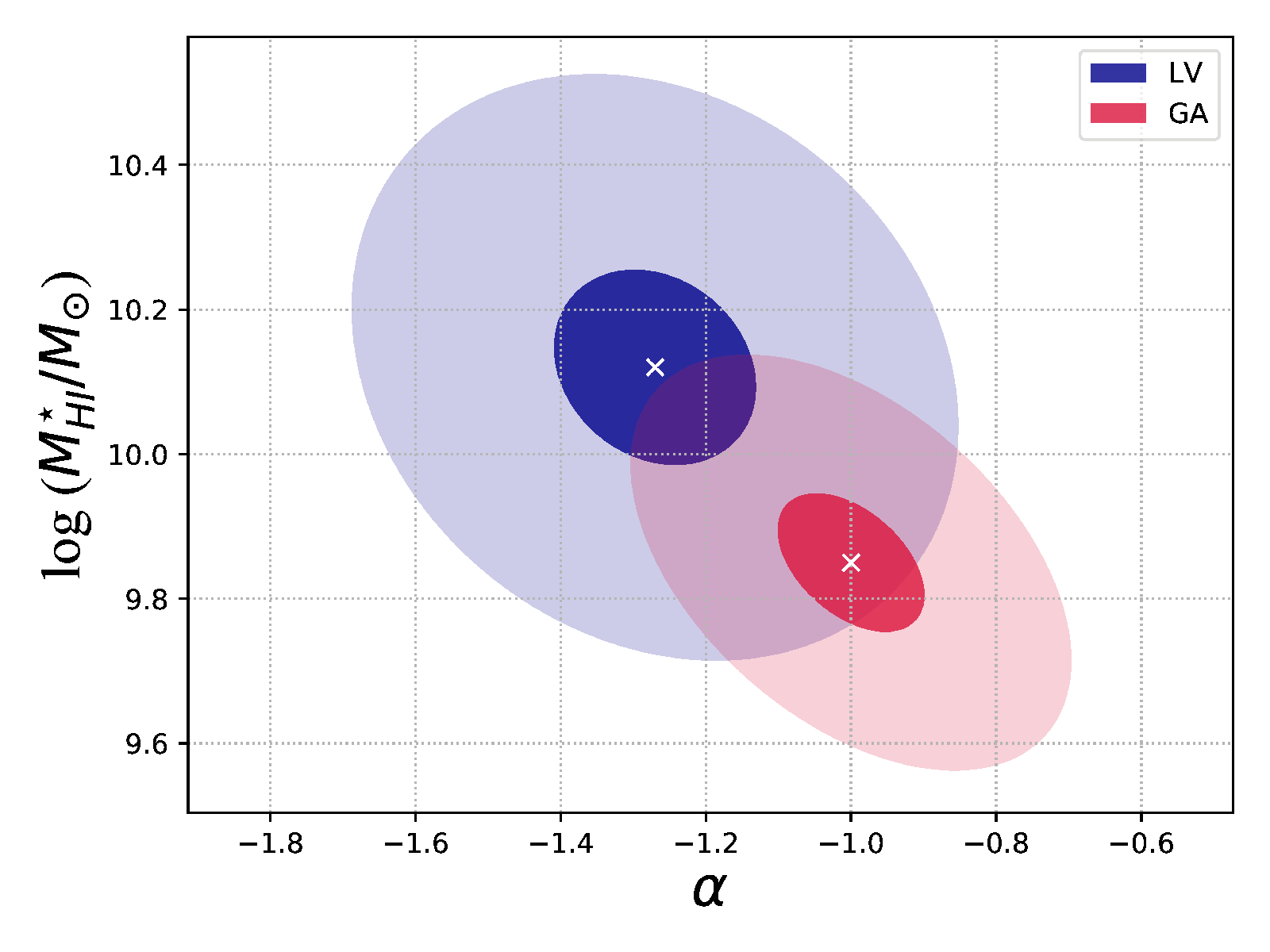}
\caption[Schechter parameters for the GA and LV]{Schechter parameters for the GA and LV samples projected on the $\alpha-\log M_{HI}^*$ plane. Ellipses are defined as in Fig. \ref{four_regions}}.
\label{GA_LV_regions}
\end{center}
\end{figure}

Confidence ellipses are shown on the $\alpha-\log M_{HI}^*$ plane for the GA and LV regions in Fig. \ref{GA_LV_regions}. The derived Schechter parameters of the GA and LV regions do not agree within the $1\sigma$ errors, confirming the previous results from the Bayesian $k$-th nearest neighbours.

We integrate the GA and LV HIMF over a mass range of $\log(M_{\rm HI}^*/M_{\odot})=8.5 - 10.8$ and compare their values with full HIZOA HIMF integrated over the same mass range. This results in an overall overdensity in the GA region of 1.35 and underdensity in the LV region of 0.59 compared to the complete HIZOA sample.

\begin{figure}
\begin{center}
\includegraphics[scale=0.5]{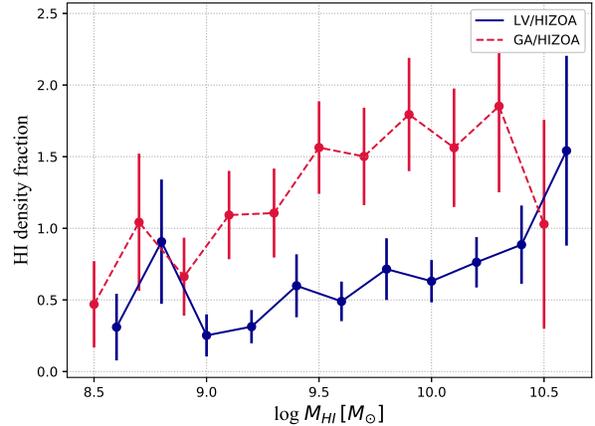}
\caption[Mass density of both GA and LV regions compared to the complete HIZOA sample]{Density fraction of both the LV (blue) and GA (red) regions compared to the complete HIZOA sample. The error bars are the Poisson errors. The LV curve is shifted horizontally by 0.1~dex for clarity.}
\label{comp_density}
\end{center}
\end{figure}

Figure \ref{comp_density} displays a more nuanced comparison of the overdensity as a function of {\HI} mass. In this Figure, we divide the derived LV and GA HIMF in Figure \ref{GA_LV_regions} by the HIZOA HIMF in Figure \ref{HIZOA_2dswml_HIMF}, resulting in the corresponding {\HI} overdensities.  
This shows that 
low-density regions appear to contain similar densities of low-mass galaxies as high-density regions, implying that low-mass galaxies are either not formed in high-density regions or (more likely) responsible for the growth of high-mass galaxies through a merging and accretion process.

\subsection{Discussion}
Analysis of the HIZOA survey finds a steeper slope at the low \HI-mass end of the HIMF in low-density regions compared to high-density regions. A simplified explanation of this result is that {\HI} gas stripping processes act on small galaxies in high-density regions, and is likely the direct results of higher tidal, collisional, and ram pressure forces in these regions. Each of these processes is important, but their exact importance in different physical situations is not straightforward to predict and, furthermore, depends on physical quantities which are difficult to measure, such as IGM density. 

The result agrees with \cite{2002ApJ...567..247R} and \cite{2005ApJ...621..215S}, who both found a similar trend of increasing $\alpha$ towards higher density regions. \cite{2001ASPC..240..507V} and \cite{2011ApJS..197...28P} also found a shallow HIMF slope for Ursa Major and six groups similar to the Local Group, respectively compare to the field HIMF. \cite{2016MNRAS.457.4393J} and \cite{2014MNRAS.444.3559M} used earlier versions of the ALFALFA catalogue  \citep{2005AJ....130.2598G,2011AJ....142..170H} and found no significant change in the low-mass slope between high and low-density regions. On the other hand, \citet{2005MNRAS.359L..30Z}  and \cite{2018MNRAS.477....2J} used the complete HIPASS and ALFALFA catalogs, respectively, and found that the low-mass slope decreases towards high-density region. Their results are opposite to the trend derived from the HIZOA data set. 

Table \ref{Comparison_all} presents a comparison of the derived HIMF parameters from a large number of different {\HI} surveys that have been used to derive the HIMF.
\begin{table*}
\begin{center}
\caption[]{Comparison of the HIMF parameters derived from different surveys. 1. \protect\cite{1997ApJ...490..173Z}, 2. \protect\cite{2001ASPC..240..507V}, 3. \protect\cite{2002ApJ...567..247R}, 4. \protect\cite{2003AJ....125.2842Z}, 5. \protect\cite{2005MNRAS.359L..30Z}, 6. \protect\cite{2005ApJ...621..215S}, 7. \protect\cite{2010ApJ...723.1359M}, 8. \protect\cite{2016MNRAS.457.4393J}, 9. \protect\cite{2018MNRAS.477....2J}, 10. \protect\cite{2017MNRAS.472.4832W}, and 11. This work}
\begin{tabular}{l c c c c c c c c c c c c c c c c}
 \hline
 \hline
Reference  &   Galaxies  & Sky Area & rms  & Mass Range & $\alpha$ & $M_{HI}^*$ & $\phi^*$ & Note on the environmental &  Survey\\
 & & (deg$^2$) & (mJy) & $\log(M_{HI}/M_{\odot})$ & & $M_{\odot}$ & Mpc$^{-3}$ & effects & \\
\hline
(1)  & 66 & 65 & 0.75 &  7.0-10.0 & $-1.20$ & 9.55 & 0.014 & --- &{\HI} Strip (AHISS)\\
(2) & 32 & --- & 0.79 &  7.0-10.0 & --- & 9.8 & 0.08 & $\rho$ $ \uparrow$   $\Rightarrow$ $\alpha$  $\uparrow$ & VLA Ursa Major\\
(3) & 265 & 430 & 3.5 & 7.0-10.0   & $-1.53$ & 9.88 & 0.005 & $\rho$ $ \uparrow$   $\Rightarrow$ $\alpha$  $\uparrow$ & Dual-Beam (ADBS)\\
(4) &  1000  & 20626   & 13  & 6.8-10.6  & $-1.30$  & 9.79 & 0.009  & --- &HIPASS (BGC) \\
(5) &  4315  & 21346  & 13  & 6.8-10.6  & $-1.37$ & 9.80 & 0.006 & $\rho$ $ \uparrow$   $\Rightarrow$ $\alpha$  $\downarrow$ & HIPASS (HICAT) \\
(6) & 2771 & 10602 & --- &  7.0-10.0  & $-1.24$ & 9.99 & 0.003 & $\rho$ $ \uparrow$   $\Rightarrow$ $\alpha$  $\uparrow$ and $M_{HI}^*$ $\downarrow$ &  Optically selected sample\\
(7) &   10119 & 2607  & 2.4  &  6.2-11.0   &  $-1.33$ & 9.96 & 0.005  &  --- & ALFALFA $40\%$\\
(8) &  20000  & 4830  & 2.4  &  6.2-11.0  & --- &  --- & --- & $\rho$ $ \uparrow$   $\Rightarrow$ $M_{HI}^*$ $\uparrow$ & ALFALFA $70\%$ \\
(9) &  ---  & 6900  & 2.4  &  6.2-11.0   & $-1.25$  & 9.94 & 0.005 & $\rho$ $ \uparrow$   $\Rightarrow$ $\alpha$  $\downarrow$ and $M_{HI}^*$ $\uparrow$& ALFALFA $100\%$ \\
(10)   &  31  & ---  & 4.0  &  6.6-9.4   & $-1.10$  & ---  & --- & $\rho$ $ \uparrow$   $\Rightarrow$ $\alpha$  $\uparrow$ &  Parkes Sculptor group\\
(11) & 751 & 1230 & 6 & 7.0-10.6  & $-1.33$ & 9.93 & 0.004 & $\rho$ $ \uparrow$   $\Rightarrow$ $\alpha$  $\uparrow$ & HIZOA\\
\hline
\end{tabular}
\label{Comparison_all}
\end{center}
\end{table*} 
This comparison shows a tension between most of the results. It is possible in some cases that the selection of different nearest neighbour metrics could cause different scales and therefore different physical processes to be probed. However, we expect these discrepancies to be resolved in the near future with forthcoming deeper and more sensitive surveys such as the Widefield ASKAP L-band Legacy All-sky Blind surveY (WALLABY; \citealt{2019MNRAS.482.3591R}), the MeerKAT Fornax Survey \citep{2016mks..confE...8S}, and the MeerKAT LADUMA survey \citep{2016mks..confE...4B}. The advantage of the WALLABY  survey will be twofold: it will be less sensitive to cosmic variance due to the large volume; and it will push the detection to lower mass dwarf galaxies with $M_{\rm HI} = 10^{8} M_{\odot}$ out to distances of 60 Mpc \citep{2012PASA...29..359K}. The MeerKAT Fornax Survey will observe the Fornax cluster with high resolution to measure the {\HI} low-mass slope down to $5\times10^5$ M$_{\odot}$, while the MeerKAT LADUMA survey \citep{2016mks..confE...4B}  will focus more on the evolution of the {\HI} mass function since it will be able to measure the {\HI} mass function out to higher redshifts. 

\section{Summary}
In this paper, we constructed the  {\HI} mass function for the Parkes {\HI} Zone of Avoidance Survey using the two-dimensional stepwise maximum likelihood method. All galaxies in the HIZOA survey that meet our selection criteria ($N=751$ galaxies) are included in the derivation of this {\HI} mass function. The average parameters of the {\HI} mass function are $\alpha = -1.33\pm0.05$, $\log(M_{HI}^*/M_{\odot})=9.93\pm0.04$, and $\phi^* = (3.9\pm0.6)\times 10^{-3}$ Mpc$^{-3}$. Comparisons of these parameters with values from the literature were made. We found that the faint-end slope derived here agrees, within the uncertainty, with HIPASS BGC, HICAT,  ALFALFA 40\%, and ALFALFA 100\%. The characteristic mass derived here agrees better, however, with the ALFALFA {\HI} mass function than with the HIPASS {\HI} mass function.  HIZOA is twice as sensitive as HIPASS, which means the depth of the survey is better matched with ALFALFA, though source confusion may be more common. 

We took advantage of the mix of high and low-density environments covered by HIZOA to test the universality of the {\HI} mass function. Two independent methods were used to define the local density environment for galaxies. The first method was the $k$-th nearest neighbour algorithm. In this method we used a Bayesian version to calculate distances to all the $k$ nearest neighbours and not only the $k$-th nearest neighbour.  This is more robust especially in the case of a sparse sample. The other method involved confining our analysis to two extreme environments located within the HIZOA survey area: the Local Void as a low-density environment and the Great Attractor as a high-density environment. 

The above mentioned two methods lead to similar conclusion that: (i) the low-mass slope $\alpha$, of the {\HI} mass function is steeper in low-density environments than in high-density environments; and (ii) there is no clear trend of the characteristic {\HI} mass $M_{\rm HI}^*$ with increasing density of the environment.

This conclusion  align closely with theoretical predictions that the stripping of the {\HI} mass occurs at higher rates in  dense regions compared to low density regions due to the increase in the interaction rate of galaxies.

Comparison with other studies shows that not all previous results are in agreement. Forthcoming {\HI} surveys with ASKAP, MeerKAT and the SKA will help resolve these issues and will be useful in understanding the details of the effect of environment on galaxy evolution.

\section*{Acknowledgments}
We acknowledge the HIZOA survey team for early access to the data. RCKK thanks the South African National Research Foundation for their support. Parts of this research were supported by the Australian Research Council Centre of Excellence for All Sky Astrophysics in 3 Dimensions (ASTRO 3D), through project number CE170100013.

\bibliographystyle{mn2e.bst}
\bibliography{references_n}

\begin{thebibliography}{}

\bibitem[\protect\citeauthoryear{{Bekki}}{{Bekki}}{1999}]{1999ApJ...510L..15B}
{Bekki} K.,  1999, \apjl, 510, L15

\bibitem[\protect\citeauthoryear{{Bekki}, {Koribalski}, {Ryder} \&
  {Couch}}{{Bekki} et~al.}{2005}]{2005MNRAS.357L..21B}
{Bekki} K.,  {Koribalski} B.~S.,  {Ryder} S.~D.,    {Couch} W.~J.,  2005,
  \mnras, 357, L21

\bibitem[\protect\citeauthoryear{{Blyth} et~al.,}{{Blyth}
  et~al.}{2016}]{2016mks..confE...4B}
{Blyth} S.  et~al., 2016, in Proceedings of MeerKAT Science: On the Pathway to
  the SKA. 25-27 May, 2016 Stellenbosch, South Africa (MeerKAT2016). p.~4

\bibitem[\protect\citeauthoryear{{Branchini} et~al.,}{{Branchini}
  et~al.}{1999}]{1999MNRAS.308....1B}
{Branchini} E.  et~al., 1999, \mnras, 308, 1

\bibitem[\protect\citeauthoryear{{Casertano} \& {Hut}}{{Casertano} \&
  {Hut}}{1985}]{1985ApJ...298...80C}
{Casertano} S.,  {Hut} P.,  1985, \apj, 298, 80

\bibitem[\protect\citeauthoryear{{Chung}, {van Gorkom}, {Kenney}, {Crowl} \&
  {Vollmer}}{{Chung} et~al.}{2009}]{2009AJ....138.1741C}
{Chung} A.,  {van Gorkom} J.~H.,  {Kenney} J.~D.~P.,  {Crowl} H.,    {Vollmer}
  B.,  2009, \aj, 138, 1741

\bibitem[\protect\citeauthoryear{{Cowie} \& {Songaila}}{{Cowie} \&
  {Songaila}}{1977}]{1977Natur.266..501C}
{Cowie} L.~L.,  {Songaila} A.,  1977, \nat, 266, 501

\bibitem[\protect\citeauthoryear{{Donley} et~al.,}{{Donley}
  et~al.}{2005}]{2005AJ....129..220D}
{Donley} J.~L.  et~al., 2005, \aj, 129, 220

\bibitem[\protect\citeauthoryear{{Dressler}}{{Dressler}}{1980}]{1980ApJ...236..351D}
{Dressler} A.,  1980, \apj, 236, 351

\bibitem[\protect\citeauthoryear{{English}, {Koribalski}, {Bland-Hawthorn},
  {Freeman} \& {McCain}}{{English} et~al.}{2010}]{2010AJ....139..102E}
{English} J.,  {Koribalski} B.,  {Bland-Hawthorn} J.,  {Freeman} K.~C.,
  {McCain} C.~F.,  2010, \aj, 139, 102

\bibitem[\protect\citeauthoryear{{Erdo{\v g}du} et~al.,}{{Erdo{\v g}du}
  et~al.}{2006}]{2006MNRAS.373...45E}
{Erdo{\v g}du} P.  et~al., 2006, \mnras, 373, 45

\bibitem[\protect\citeauthoryear{{Giovanelli} et~al.,}{{Giovanelli}
  et~al.}{2005}]{2005AJ....130.2598G}
{Giovanelli} R.  et~al., 2005, \aj, 130, 2598

\bibitem[\protect\citeauthoryear{{Gott} III \& {Gunn}}{{Gott} \&
  {Gunn}}{1971}]{1971ApJ...169L..13G}
{Gott} III J.~R.,  {Gunn} J.~E.,  1971, \apjl, 169, L13

\bibitem[\protect\citeauthoryear{{Gunn} \& {Gott} III}{{Gunn} \&
  {Gott}}{1972}]{1972ApJ...176....1G}
{Gunn} J.~E.,  {Gott} III J.~R.,  1972, \apj, 176, 1

\bibitem[\protect\citeauthoryear{{Haynes} et~al.,}{{Haynes}
  et~al.}{2011}]{2011AJ....142..170H}
{Haynes} M.~P.  et~al., 2011, \aj, 142, 170

\bibitem[\protect\citeauthoryear{{Ivezi{\'c}}, {Vivas}, {Lupton} \&
  {Zinn}}{{Ivezi{\'c}} et~al.}{2005}]{2005AJ....129.1096I}
{Ivezi{\'c}} {\v Z}.,  {Vivas} A.~K.,  {Lupton} R.~H.,    {Zinn} R.,  2005,
  \aj, 129, 1096

\bibitem[\protect\citeauthoryear{{Jones}, {Haynes}, {Giovanelli} \&
  {Moorman}}{{Jones} et~al.}{2018}]{2018MNRAS.477....2J}
{Jones} M.~G.,  {Haynes} M.~P.,  {Giovanelli} R.,    {Moorman} C.,  2018,
  \mnras, 477, 2

\bibitem[\protect\citeauthoryear{{Jones}, {Papastergis}, {Haynes} \&
  {Giovanelli}}{{Jones} et~al.}{2016}]{2016MNRAS.457.4393J}
{Jones} M.~G.,  {Papastergis} E.,  {Haynes} M.~P.,    {Giovanelli} R.,  2016,
  \mnras, 457, 4393

\bibitem[\protect\citeauthoryear{{Koribalski}}{{Koribalski}}{2012}]{2012PASA...29..359K}
{Koribalski} B.~S.,  2012, pasa, 29, 359

\bibitem[\protect\citeauthoryear{{Koribalski} et~al.,}{{Koribalski}
  et~al.}{2004}]{2004AJ....128...16K}
{Koribalski} B.~S.  et~al., 2004, \aj, 128, 16

\bibitem[\protect\citeauthoryear{{Kraan-Korteweg}, {Cluver}, {Bilicki},
  {Jarrett}, {Colless}, {Elagali}, {B{\"o}hringer} \& {Chon}}{{Kraan-Korteweg}
  et~al.}{2017}]{2017MNRAS.466L..29K}
{Kraan-Korteweg} R.~C.,  {Cluver} M.~E.,  {Bilicki} M.,  {Jarrett} T.~H.,
  {Colless} M.,  {Elagali} A.,  {B{\"o}hringer} H.,    {Chon} G.,  2017,
  \mnras, 466, L29

\bibitem[\protect\citeauthoryear{{Kraan-Korteweg}, {Shafi}, {Koribalski},
  {Staveley-Smith}, {Buckland}, {Henning} \& {Fairall}}{{Kraan-Korteweg}
  et~al.}{2008}]{2008glv..book...13K}
{Kraan-Korteweg} R.~C.,  {Shafi} N.,  {Koribalski} B.~S.,  {Staveley-Smith} L.,
   {Buckland} P.,  {Henning} P.~A.,    {Fairall} A.~P.,  2008, {Outlining the
  Local Void with the Parkes HI ZOA and Galactic Bulge Surveys}.
p.~13

\bibitem[\protect\citeauthoryear{{Lavaux} \& {Hudson}}{{Lavaux} \&
  {Hudson}}{2011}]{2011MNRAS.416.2840L}
{Lavaux} G.,  {Hudson} M.~J.,  2011, \mnras, 416, 2840

\bibitem[\protect\citeauthoryear{{Loveday}}{{Loveday}}{2000}]{2000MNRAS.312..557L}
{Loveday} J.,  2000, \mnras, 312, 557

\bibitem[\protect\citeauthoryear{{Lynden-Bell}, {Faber}, {Burstein}, {Davies},
  {Dressler}, {Terlevich} \& {Wegner}}{{Lynden-Bell}
  et~al.}{1988}]{1988ApJ...326...19L}
{Lynden-Bell} D.,  {Faber} S.~M.,  {Burstein} D.,  {Davies} R.~L.,  {Dressler}
  A.,  {Terlevich} R.~J.,    {Wegner} G.,  1988, \apj, 326, 19

\bibitem[\protect\citeauthoryear{{Martin}, {Papastergis}, {Giovanelli},
  {Haynes}, {Springob} \& {Stierwalt}}{{Martin}
  et~al.}{2010}]{2010ApJ...723.1359M}
{Martin} A.~M.,  {Papastergis} E.,  {Giovanelli} R.,  {Haynes} M.~P.,
  {Springob} C.~M.,    {Stierwalt} S.,  2010, \apj, 723, 1359

\bibitem[\protect\citeauthoryear{{Meekins}, {Fritz}, {Chubb} \&
  {Friedman}}{{Meekins} et~al.}{1971}]{1971Natur.231..107M}
{Meekins} J.~F.,  {Fritz} G.,  {Chubb} T.~A.,    {Friedman} H.,  1971, \nat,
  231, 107

\bibitem[\protect\citeauthoryear{{Meyer} et~al.,}{{Meyer}
  et~al.}{2004}]{2004MNRAS.350.1195M}
{Meyer} M.~J.  et~al., 2004, \mnras, 350, 1195

\bibitem[\protect\citeauthoryear{{Moorman}, {Vogeley}, {Hoyle}, {Pan}, {Haynes}
  \& {Giovanelli}}{{Moorman} et~al.}{2014}]{2014MNRAS.444.3559M}
{Moorman} C.~M.,  {Vogeley} M.~S.,  {Hoyle} F.,  {Pan} D.~C.,  {Haynes} M.~P.,
    {Giovanelli} R.,  2014, \mnras, 444, 3559

\bibitem[\protect\citeauthoryear{{Obreschkow}, {Kl{\"o}ckner}, {Heywood},
  {Levrier} \& {Rawlings}}{{Obreschkow} et~al.}{2009}]{2009ApJ...703.1890O}
{Obreschkow} D.,  {Kl{\"o}ckner} H.-R.,  {Heywood} I.,  {Levrier} F.,
  {Rawlings} S.,  2009, \apj, 703, 1890

\bibitem[\protect\citeauthoryear{{Papastergis}, {Martin}, {Giovanelli} \&
  {Haynes}}{{Papastergis} et~al.}{2011}]{2011ApJ...739...38P}
{Papastergis} E.,  {Martin} A.~M.,  {Giovanelli} R.,    {Haynes} M.~P.,  2011,
  \apj, 739, 38

\bibitem[\protect\citeauthoryear{{Pisano}, {Barnes}, {Staveley-Smith},
  {Gibson}, {Kilborn} \& {Freeman}}{{Pisano}
  et~al.}{2011}]{2011ApJS..197...28P}
{Pisano} D.~J.,  {Barnes} D.~G.,  {Staveley-Smith} L.,  {Gibson} B.~K.,
  {Kilborn} V.~A.,    {Freeman} K.~C.,  2011, \apjs, 197, 28

\bibitem[\protect\citeauthoryear{{Rauzy}}{{Rauzy}}{2001}]{2001MNRAS.324...51R}
{Rauzy} S.,  2001, \mnras, 324, 51

\bibitem[\protect\citeauthoryear{{Reynolds} et~al.,}{{Reynolds}
  et~al.}{2019}]{2019MNRAS.482.3591R}
{Reynolds} T.~N.  et~al., 2019, \mnras, 482, 3591

\bibitem[\protect\citeauthoryear{{Rosenberg} \& {Schneider}}{{Rosenberg} \&
  {Schneider}}{2002}]{2002ApJ...567..247R}
{Rosenberg} J.~L.,  {Schneider} S.~E.,  2002, \apj, 567, 247

\bibitem[\protect\citeauthoryear{{Schechter}}{{Schechter}}{1976}]{1976ApJ...203..297S}
{Schechter} P.,  1976, \apj, 203, 297

\bibitem[\protect\citeauthoryear{{Schmidt}}{{Schmidt}}{1968}]{1968ApJ...151..393S}
{Schmidt} M.,  1968, \apj, 151, 393

\bibitem[\protect\citeauthoryear{{Serra} et~al.,}{{Serra}
  et~al.}{2016}]{2016mks..confE...8S}
{Serra} P.  et~al., 2016, in Proceedings of MeerKAT Science: On the Pathway to
  the SKA. 25-27 May, 2016 Stellenbosch, South Africa (MeerKAT2016). p.~8

\bibitem[\protect\citeauthoryear{{Serra} et~al.,}{{Serra}
  et~al.}{2013}]{2013MNRAS.428..370S}
{Serra} P.  et~al., 2013, \mnras, 428, 370

\bibitem[\protect\citeauthoryear{{Serra} et~al.,}{{Serra}
  et~al.}{2012}]{2012MNRAS.422.1835S}
{Serra} P.  et~al., 2012, \mnras, 422, 1835

\bibitem[\protect\citeauthoryear{{Shostak}, {Sullivan} III \&
  {Allen}}{{Shostak} et~al.}{1984}]{1984A&A...139...15S}
{Shostak} G.~S.,  {Sullivan} III W.~T.,    {Allen} R.~J.,  1984, \aap, 139, 15

\bibitem[\protect\citeauthoryear{{Sorce}, {Colless}, {Kraan-Korteweg} \&
  {Gottl{\"o}ber}}{{Sorce} et~al.}{2017}]{2017MNRAS.471.3087S}
{Sorce} J.~G.,  {Colless} M.,  {Kraan-Korteweg} R.~C.,    {Gottl{\"o}ber} S.,
  2017, \mnras, 471, 3087

\bibitem[\protect\citeauthoryear{{Springel} et~al.,}{{Springel}
  et~al.}{2005}]{2005Natur.435..629S}
{Springel} V.  et~al., 2005, \nat, 435, 629

\bibitem[\protect\citeauthoryear{{Springob}, {Haynes} \&
  {Giovanelli}}{{Springob} et~al.}{2005}]{2005ApJ...621..215S}
{Springob} C.~M.,  {Haynes} M.~P.,    {Giovanelli} R.,  2005, \apj, 621, 215

\bibitem[\protect\citeauthoryear{{Staveley-Smith}, {Kraan-Korteweg},
  {Schr{\"o}der}, {Henning}, {Koribalski}, {Stewart} \&
  {Heald}}{{Staveley-Smith} et~al.}{2016}]{2016AJ....151...52S}
{Staveley-Smith} L.,  {Kraan-Korteweg} R.~C.,  {Schr{\"o}der} A.~C.,  {Henning}
  P.~A.,  {Koribalski} B.~S.,  {Stewart} I.~M.,    {Heald} G.,  2016, \aj, 151,
  52

\bibitem[\protect\citeauthoryear{{Tully} \& {Fisher}}{{Tully} \&
  {Fisher}}{1987}]{1987ang..book.....T}
{Tully} R.~B.,  {Fisher} J.~R.,  1987, {Atlas of Nearby Galaxies}

\bibitem[\protect\citeauthoryear{{Verheijen}, {Trentham}, {Tully} \&
  {Zwaan}}{{Verheijen} et~al.}{2001}]{2001ASPC..240..507V}
{Verheijen} M.~A.~W.,  {Trentham} N.,  {Tully} B.,    {Zwaan} M.,  2001, in
  {Hibbard} J.~E.,  {Rupen} M.,   {van Gorkom} J.~H.,  eds,  Astronomical
  Society of the Pacific Conference Series Vol. 240, Gas and Galaxy Evolution.
  p.~507

\bibitem[\protect\citeauthoryear{{Westmeier}, {Koribalski} \&
  {Braun}}{{Westmeier} et~al.}{2013}]{2013MNRAS.434.3511W}
{Westmeier} T.,  {Koribalski} B.~S.,    {Braun} R.,  2013, \mnras, 434, 3511

\bibitem[\protect\citeauthoryear{{Westmeier} et~al.,}{{Westmeier}
  et~al.}{2017}]{2017MNRAS.472.4832W}
{Westmeier} T.  et~al., 2017, \mnras, 472, 4832

\bibitem[\protect\citeauthoryear{{Woudt}, {Kraan-Korteweg} \&
  {Fairall}}{{Woudt} et~al.}{1999}]{1999A&A...352...39W}
{Woudt} P.~A.,  {Kraan-Korteweg} R.~C.,    {Fairall} A.~P.,  1999, \aap, 352,
  39

\bibitem[\protect\citeauthoryear{{Zwaan}, {Briggs}, {Sprayberry} \&
  {Sorar}}{{Zwaan} et~al.}{1997}]{1997ApJ...490..173Z}
{Zwaan} M.~A.,  {Briggs} F.~H.,  {Sprayberry} D.,    {Sorar} E.,  1997, \apj,
  490, 173

\bibitem[\protect\citeauthoryear{{Zwaan}, {Meyer}, {Staveley-Smith} \&
  {Webster}}{{Zwaan} et~al.}{2005}]{2005MNRAS.359L..30Z}
{Zwaan} M.~A.,  {Meyer} M.~J.,  {Staveley-Smith} L.,    {Webster} R.~L.,  2005,
  \mnras, 359, L30

\bibitem[\protect\citeauthoryear{{Zwaan} et~al.,}{{Zwaan}
  et~al.}{2003}]{2003AJ....125.2842Z}
{Zwaan} M.~A.  et~al., 2003, \aj, 125, 2842

\end{thebibliography}

\appendix
\section{Applying the classic $k$-th nearest neighbour method}
\begin{figure}
\begin{center}
\includegraphics[scale=0.52]{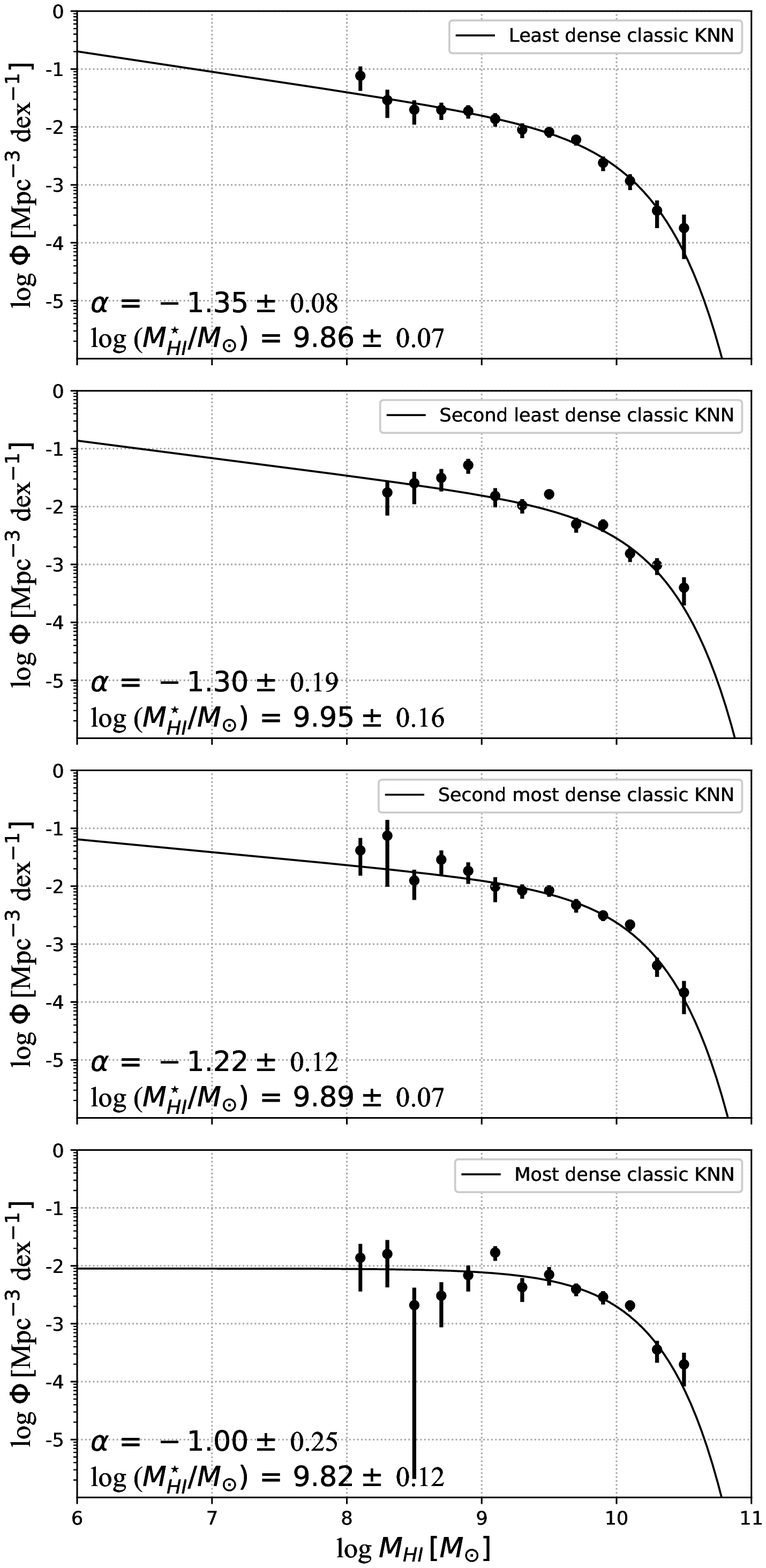}
\caption[Four HIZOA]{Four HIZOA {\HI} mass function derived via the 2DSWML method based on the classic density estimation used in HIPASS. For comparison, the HIMFs are offset vertically to the normalization of the whole sample HIMF in Fig. \ref{HIZOA_2dswml_HIMF}.}
\label{compare_density_Alpha_Mstar_classic}
\end{center}
\end{figure}

There is a strong argument for applying the HIPASS method of defining the local density on the HIZOA data to apply a direct comparison. In this appendix, we used the classic $k$-th nearest neighbour algorithm that was used by \cite{2005MNRAS.359L..30Z} to define the environment. We divided our sample into four sub-samples and derived low-mass slope and characteristic mass for each sub-sample. Figure \ref{compare_density_Alpha_Mstar_classic} shows the HIMF for these four sub-samples ranging from least dense environment (top-panel) to most dense environment (bottom-panel). 

Table \ref{Schechter_P_C} presents the derived Schechter parameters for these four sub-sample ranging from low to high-density environment.

\begin{table}
\begin{center}
\caption[Schechter parameters for the HIMF]{Schechter parameters for four sub-samples divided according to their local density derived by the classic version of the $k$-th nearest neighbour algorithm.}
\begin{tabular}{l c c}
\hline
\hline
Region  &   $\alpha$ & $\log(M_{HI}^*/M_{\odot})$\\
\hline
Least dense &   $-1.35\pm0.08$ &  $9.86\pm0.07$  \\
Second least dense &   $-1.30\pm0.19$ &  $9.95\pm0.16$ \\
Second most dense &   $-1.22\pm0.12$ &  $9.89\pm0.07$ \\
Most dense &   $-1.00\pm0.25$ &  $9.82\pm0.12$ \\
\hline
\end{tabular}
\label{Schechter_P_C}
\end{center}
\end{table} 

The exact same trend for the low mass slope $\alpha$ that is found when using the Bayesian $k$ nearest neighbours analysis is also found when using the classic $k$-th nearest neighbour. 

\section{Correction of the Bayesian $k$ nearest neighbours $\Sigma_6$}
\label{BBB}
Because HIZOA is a flux-limited sample, the number of galaxies eventually starts dropping as a function of redshift. This implies that galaxies will appear increasingly isolated at larger redshift. If this is not taken into account, different density samples could mistakenly represent different distance ranges.

Whereas previous authors \citep{2005MNRAS.359L..30Z} have fit the data to correct for this bias, we chose a more robust correction based on the $S^3-SAX$ semi-analytic simulation \citep{2009ApJ...703.1890O}. This simulation is based, in turn, on the the Millennium simulation of cosmic structure \citep{2005Natur.435..629S}. Two samples have been selected: (A) a volume limited sample with $\log(M_{HI}^*/M_{\odot}) > 8.0$ and $v_{hel} < 12000$ km s$^{-1}$; (B) a flux-limited sample with the same hybrid-limit as HIZOA sample and $v_{hel} < 12000$ km s$^{-1}$. For each sample, we used the same Bayesian $k$ nearest neighbours algorithm to calculate the sixth nearest neighbour density, $\Sigma_6$. We will refer to the sixth nearest neighbour density derived from the volume-limited sample as real, $\Sigma_6^{real}$, while for the flux-limited sample we will call it measured, $\Sigma_6^{measured}$.

\begin{figure}
\begin{center}
\includegraphics[scale=0.42]{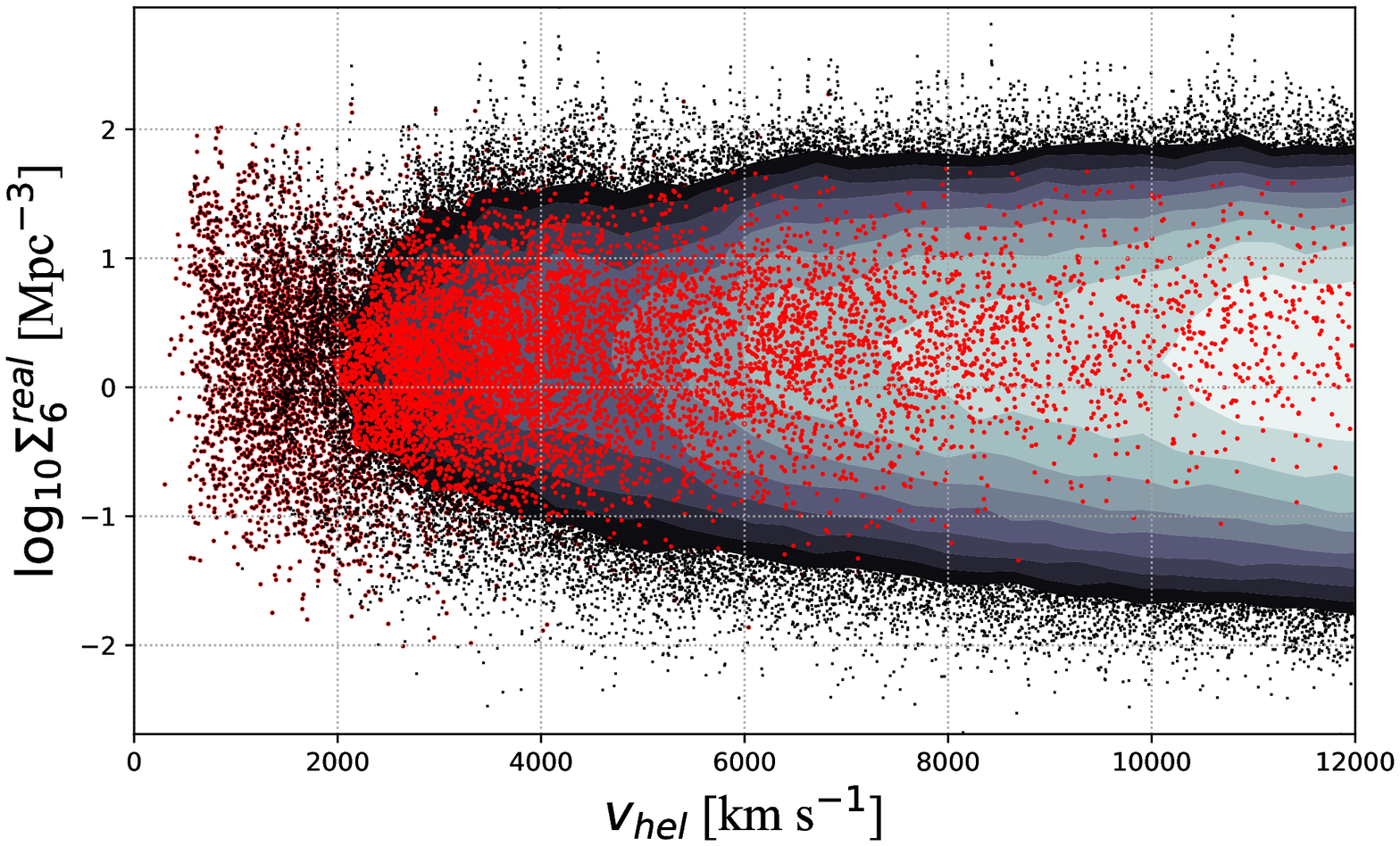}
\caption[Scatter plot of the sixth nearest neighbour density, $\Sigma_6^{real}$, with contours over-dense regions using the Bayesian $k$ nearest neighbours algorithm for a volume-limited sample]{Scatter plot of the sixth nearest neighbour density, $\Sigma_6$, with contours over-dense regions using the Bayesian $k$ nearest neighbours algorithm for a volume-limited sample derived from the $S^3-SAX$ simulation. The red dots highlight the flux-limited sample also derived from the $S^3-SAX$ simulation.}
\label{rho_vol_for_flux}
\end{center}
\end{figure}

Figure \ref{rho_vol_for_flux} shows the distribution of $\Sigma_6^{real}$ as a function of redshift. As expected, for a volume-limited sample there is no obvious trend of $\Sigma_6^{real}$ with redshift. We highlighted the flux-limited sample by the red dots.

\begin{figure}
\begin{center}
\includegraphics[scale=0.52]{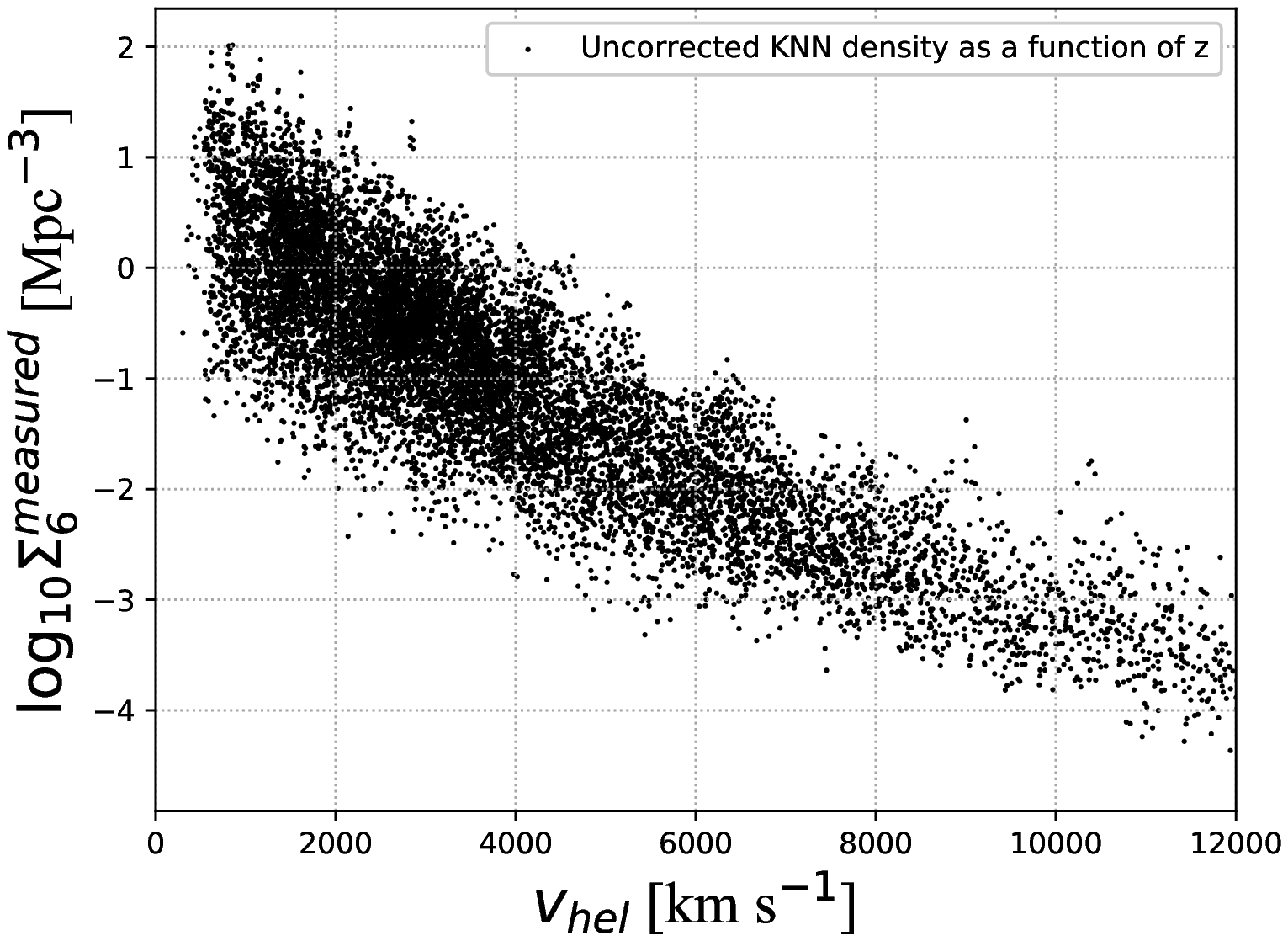}
\caption[Scatter plot of the sixth nearest neighbour density, $\Sigma_6^{measured}$, using the Bayesian $k$ nearest neighbours algorithm for a flux-limited sample]{Scatter plot of the sixth nearest neighbour density, $\Sigma_6$, using the Bayesian $k$ nearest neighbours algorithm for a flux-limited sample derived from the $S^3-SAX$ simulation.}
\label{rho_uncorrected_flux_vs_redshift}
\end{center}
\end{figure}

On the contrary, Fig. \ref{rho_uncorrected_flux_vs_redshift} shows the distribution of $\Sigma_6^{measured}$ as a function of redshift. The same trend of decreasing $\Sigma_6^{measured}$ with redshift is seen as in the HIZOA sample in the top-panel of Fig. \ref{uncor_vs_cor}.

\begin{figure}
\begin{center}
\includegraphics[scale=0.52]{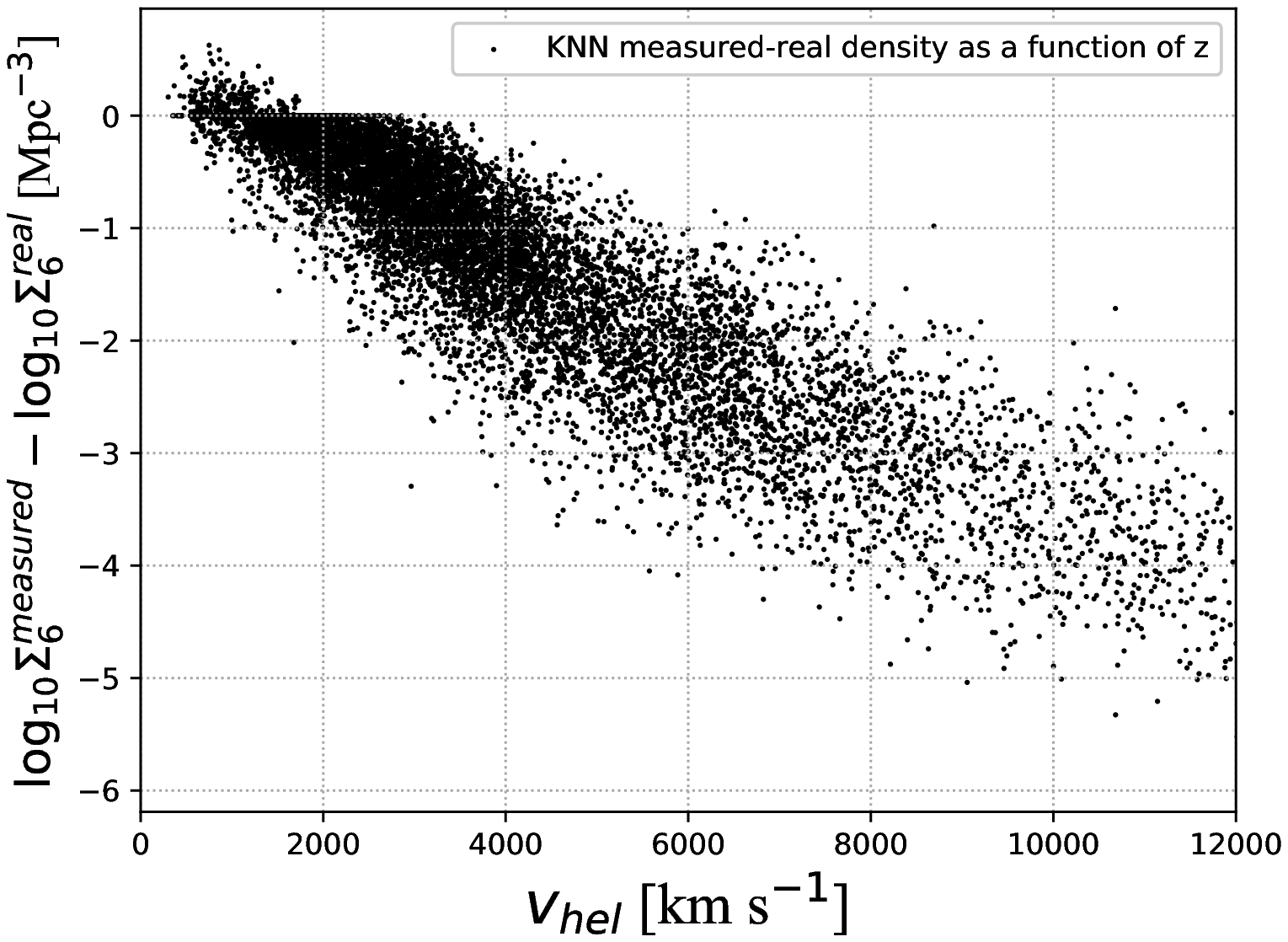}
\caption[Scatter plot of the log of the ratio measured$/$real density as a function of redshift for a flux-limited sample]{Scatter plot of the log of the ratio measured$/$real density as a function of redshift for a flux-limited sample derived from the $S^3-SAX$ simulation.}
\label{rho_ratio_mesured_real_vs_redshift}
\end{center}
\end{figure}

To extract the exact trend that we should use to correct for our sample, we calculate the log of the ratio, $\Sigma_6^{measured}/\Sigma_6^{real}$, and plot the distribution in Fig. \ref{rho_ratio_mesured_real_vs_redshift}.

We fit a power-law to the mean and rms of the log of the ratio ($\Sigma_6^{measured}/\Sigma_6^{real}$) in 1000 km/s bins. We then subtract this derived power-law from the HIZOA $\Sigma_6^{uncorrected}$ in the top-panel of Fig. \ref{uncor_vs_cor} to obtain the corrected $\Sigma_6^{corrected}$ shown in the bottom-panel of Fig. \ref{uncor_vs_cor}.

\begin{figure}
\begin{center}
\includegraphics[scale=0.52]{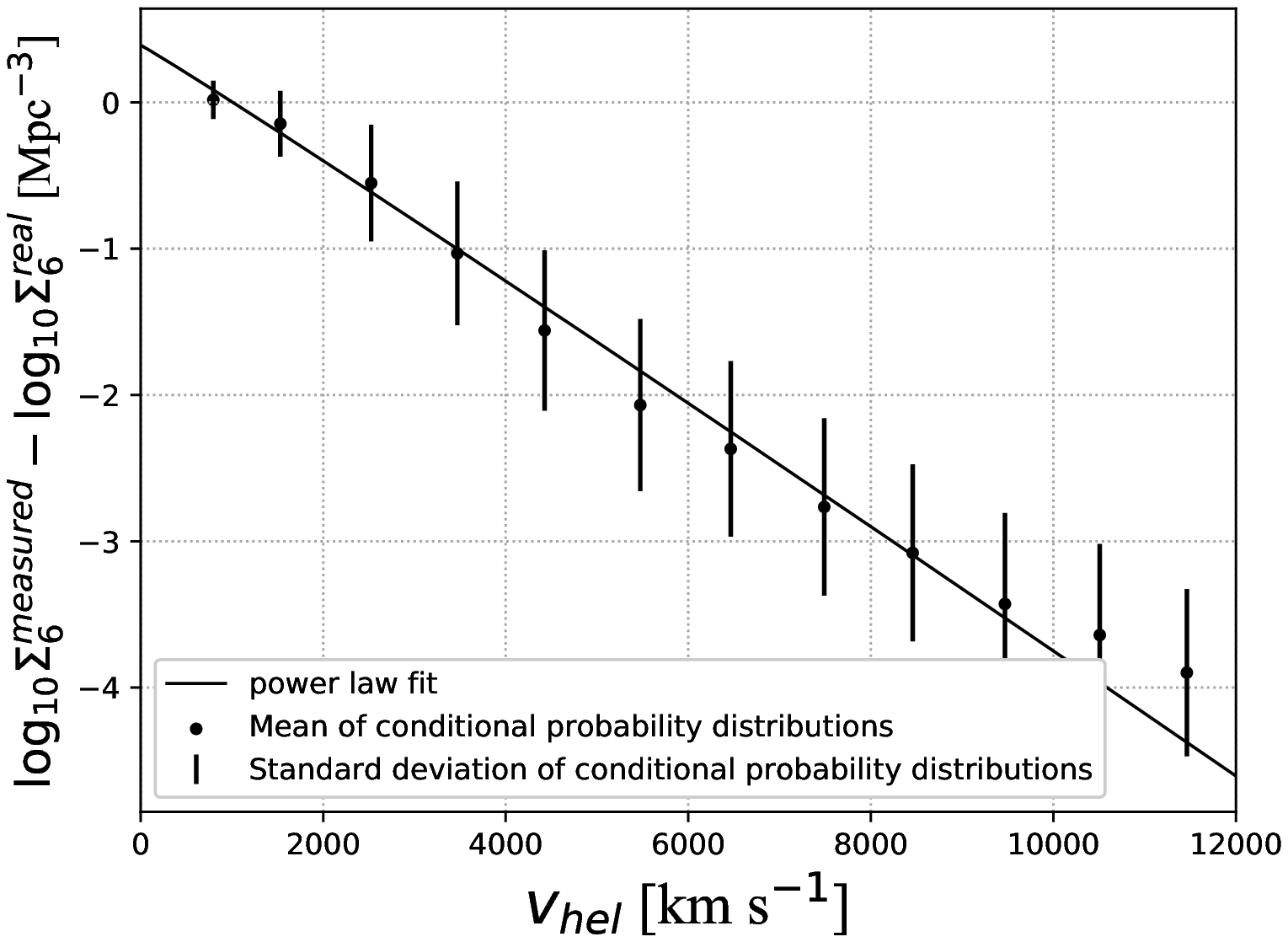}
\caption[The conditional probability distribution of the ratio measured$/$real density as a function of redshift for a flux-limited sample]{Mean and standard deviation of the conditional probability distribution of the ratio measured$/$real density as a function of redshift for a flux-limited sample. The solid line presents the power-law fit tot the conditional probability distribution.}
\label{power_law_fit_measured_real}
\end{center}
\end{figure}

The exact same method has been applied to correct the classic $k$-th nearest neighbour density but calculating classic $\Sigma_6$ for both volume- and flux-limited samples.

\label{lastpage}

\bsp

\end{document}